\documentclass[aps,prb,reprint,superscriptaddress,floatfix,longbibliography]{revtex4-2}

\usepackage{amsmath}
\usepackage{booktabs}
\usepackage{graphicx}
\usepackage{siunitx}
\usepackage{xcolor}
\usepackage[colorlinks=true,citecolor=blue,linkcolor=blue,urlcolor=blue]{hyperref}

\graphicspath{{Figures/}}
\begin{document}

\title{Contrasting anisotropic lattice responses of the $5d^1$ wolframites MgReO$_4$ and ZnReO$_4$}

\author{Ugne~Miniotaite}
\email{ugnem@kth.se}
\affiliation{Department of Applied Physics, KTH Royal Institute of Technology, SE-106 91 Stockholm, Sweden}

\author{Hiroya~Sakurai}
\affiliation{National Institute for Materials Science (NIMS), Namiki, Tsukuba, Ibaraki, Japan}

\author{Jun~Sugiyama}
\affiliation{Neutron Science and Technology Center, Comprehensive Research Organization for Science and Society (CROSS), Tokai, Japan}

\author{Toru~Ishigaki}
\affiliation{Neutron Industrial Application Promotion Center, Comprehensive Research Organization for Science and Society (CROSS), Tokai, Japan}

\author{Martin~M{\aa}nsson}
\email{condmat@kth.se}
\affiliation{Department of Applied Physics, KTH Royal Institute of Technology, SE-106 91 Stockholm, Sweden}

\author{Yasmine~Sassa}
\email{sassa@kth.se}
\affiliation{Department of Applied Physics, KTH Royal Institute of Technology, SE-106 91 Stockholm, Sweden}

\date{\today}

\begin{abstract}
Spin-orbit-entangled $5d^1$ oxides can exhibit lattice responses associated with coupled structural, orbital, and magnetic degrees of freedom. We use temperature-dependent time-of-flight neutron powder diffraction (NPD) to investigate the monoclinic wolframite-type rhenium oxides MgReO$_4$ and ZnReO$_4$ between 7 and 300~K. Both compounds retain $P2/c$ symmetry throughout the measured range, with no additional Bragg reflections or peak splitting resolved. However, their thermal expansion differs. ZnReO$_4$ exhibits smooth, monotonic anisotropic expansion, whereas MgReO$_4$ develops a broad plateau in the $b$ lattice parameter and a slight contraction along $c$, producing a shallow minimum, over approximately 200--280~K. The $a$ lattice parameter and unit-cell volume continue to increase monotonically. Independent Le Bail fits to the high-resolution backscattering-bank data reproduce the plateau and minimum. The anomalous lattice response extends toward the previously reported susceptibility and heat-capacity features near $T^*\simeq275$--280~K, although the $c$-axis minimum occurs near 250~K. Whether the lattice response and the bulk features share a microscopic origin remains open; anharmonic lattice dynamics, local or isostructural rearrangements, and magnetoelastic coupling to short-range correlations remain possible explanations.
\end{abstract}

\maketitle

\section{Introduction}
In $5d^1$ transition-metal oxides, electron--electron correlations and spin-orbit coupling (SOC) can occur on comparable energy scales. Their interplay with crystal-field effects can give rise to strongly entangled spin and orbital degrees of freedom \cite{Jackeli2009, WitczakKrempa2014, Takayama2021}, allowing these systems to exhibit a wide range of unconventional electronic and magnetic phases, including Mott insulating phases \cite{Erickson2007}, proposed quantum spin-liquid states \cite{Chen2010}, and multipolar ordered states \cite{Takayama2021, Pourovskii2025}.

The wolframite-type rhenium oxides MgReO$_4$ and ZnReO$_4$ provide a low-symmetry setting in which to examine lattice--electronic coupling. Both crystallize in monoclinic $P2/c$ symmetry and contain distorted, edge-sharing ReO$_6$ octahedra arranged in zigzag chains along the $c$ axis~\cite{Sleight_1975,ugne_AReO4}. Muon-spin spectroscopy ($\mu^+$SR) and bulk susceptibility identify magnetic ordering near $T_{\mathrm N}=83$~K for MgReO$_4$ and 126~K for ZnReO$_4$. First-principles-based calculations predict predominantly octupolar order with a small ordered dipolar component in both compounds, while no magnetic Bragg contribution has been resolved in neutron powder diffraction (NPD)~\cite{ugne_AReO4}.

An additional high-temperature feature was found in MgReO$_4$: the magnetic susceptibility changes slope near 280~K and the heat capacity shows a weak anomaly near 275~K, without thermal hysteresis~\cite{Nocerino_MgReO4}. We use $T^*\simeq275$--280~K (labeled $T_{\mathrm{C2}}$ in the earlier work \cite{Nocerino_MgReO4}) for this temperature range. Because no static magnetic order was detected by $\mu^+$SR above $T_{\mathrm N}$, a structural transition was proposed as a possible origin of this feature~\cite{Nocerino_MgReO4}. Whether $T^*$ is accompanied by a change in crystallographic symmetry, reflects a subtler isostructural lattice response, or is associated with local correlations or dynamics remains unresolved.

Here, we compare the average crystal structures of MgReO$_4$ and ZnReO$_4$ from 7 to 300~K using temperature-dependent time-of-flight NPD. Within the resolution of our NPD measurements, the diffraction patterns of both compounds remain with the average $P2/c$ structure within the entire temperature range. However, we find contrasting thermal expansion behaviors. ZnReO$_4$ expands monotonically along all three axes. In MgReO$_4$, $b$ plateaus and $c$ reaches a shallow minimum near 250~K while the unit-cell volume continues to increase. The broader lattice response in MgReO$_4$ extends toward $T^*$.

\section{Experimental and analysis methods}
Polycrystalline MgReO$_4$ and ZnReO$_4$ were synthesized by high-pressure synthesis at 6~GPa and \SI{1300}{\celsius}, following the procedure previously reported~\cite{ugne_AReO4}. Because the materials are sensitive to moisture, sample preparation and loading were performed under an inert atmosphere in a glove box. Approximately \SI{0.5}{\gram} of each powder was sealed in a cylindrical vanadium container with an internal diameter of \SI{5}{\milli\meter} using indium wire and aluminum screws.

Time-of-flight NPD measurements were performed on the iMATERIA diffractometer at J-PARC, Japan~\cite{Ishigaki2009}. Data from the high-resolution backscattering (BS) detector bank, with $\Delta d/d\simeq0.15\%$, and the special-environment (SE) detector bank, centered around $2\theta=90^\circ$, were used for the analyses described below. Regions containing unidentified impurity reflections were excluded from the refinements. Excluded regions and residuals are shown in the Supplemental
Material (SM)~\cite{supplemental_reo4}.

For ZnReO$_4$, diffraction data were acquired continuously during heating at an average rate of \SI{1}{\kelvin\per\minute}. MgReO$_4$ was measured during cooling with the same settings. However, its actual cooling rate was approximately 0.4--\SI{0.8}{\kelvin\per\minute} between 300 and 40~K and increased below 40~K.We integrated the data in 20-minute slices and assigned each slice its mean temperature (Tables~S1 and~S2 of the SM). Additional high-statistics patterns acquired at selected temperatures were analyzed separately and used as a qualitative check on the lattice parameters obtained from the time-sliced refinements.

Rietveld refinement was performed on the BS-bank data for both compounds using the FullProf Suite~\cite{RodriguezCarvajal1993}. A consistent profile and structural model was applied across each temperature series. The lattice parameters obtained from the time-sliced refinements were compared with those from static high-statistics measurements. Representative refinement profiles and the corresponding residuals are presented in the SM.

Independent Le Bail fits were performed for MgReO$_4$ using both the BS- and SE-bank data to check the temperature-dependent lattice trends obtained from the Rietveld refinements of the time-sliced data. These profile fits determine the lattice parameters without refining atomic coordinates or isotropic displacement parameters. Detector-bank-dependent offsets occur in the absolute lattice parameters, particularly for $c$. We therefore compare relative changes within each refinement series. The BS-bank Le Bail fits provide the most direct comparison with the Rietveld results.

The relative changes in the lattice parameters, monoclinic angle, and unit-cell volume are defined as
\begin{equation}
 \Delta x(T)=100\,\frac{x(T)-x(T_0)}{x(T_0)},
 \label{eq:relative_change}
\end{equation}
where $x=a,b,c,\beta,V$, and $T_0\simeq7$~K.
For the monoclinic cell, the unit-cell volume is
$V=abc\sin\beta$. For each compound, the time-sliced and static high-statistics results were referenced to the values obtained from the same 7~K high-statistics refinement.

\section{Results}

\subsection{Crystal structure}

\begin{figure*}[t]
 \centering
 \includegraphics[width=0.86\textwidth]{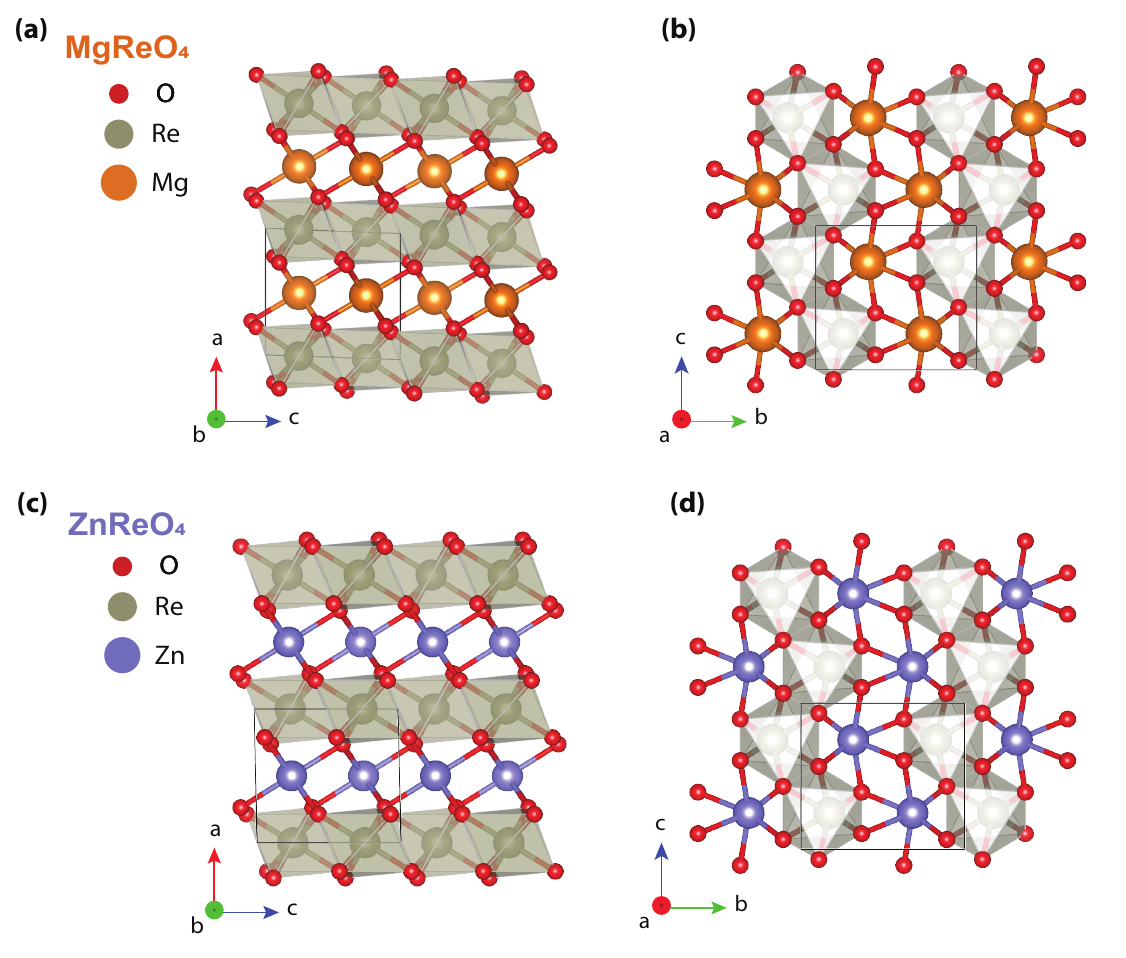}
\caption{Monoclinic wolframite-type structures of (a,b) MgReO$_4$ and (c,d) ZnReO$_4$ in space group $P2/c$. Panels (a,c) and (b,d) show projections along the $b$ and $a$ axes, respectively. Shaded polyhedra represent distorted ReO$_6$ octahedra, which share edges to form zigzag chains along $c$. The chains are separated by Mg or Zn cations. Atom colors are indicated in the legends. The structures were visualized using VESTA~\cite{VESTA}.}
 \label{fig:structure}
\end{figure*}

Both MgReO$_4$ and ZnReO$_4$ adopt the monoclinic wolframite-type structure (space group $P2/c$), previously described by Sleight \textit{et al.}~\cite{Sleight_1975} and characterized in our earlier study~\cite{ugne_AReO4}. The structure is characterized by distorted, edge-sharing ReO$_6$ octahedra forming zigzag chains along the $c$ axis, separated by Mg or Zn coordination polyhedra along the $a$ axis (see Fig.~\ref{fig:structure}).

Our temperature-dependent NPD shows that this structure remains stable between 7 and 300~K. We observe no additional Bragg reflections or peak splitting throughout the measured range (see Fig.~S2 of the SM), and the structural refinements show no change of the $P2/c$ symmetry in either compound. Thus, within the resolution of the present NPD measurements, we find no evidence of a long-range symmetry-lowering structural transition over the investigated temperature range.

\subsection{Contrasting lattice responses}

\begin{figure*}[t]
 \centering
 \includegraphics[width=\textwidth]{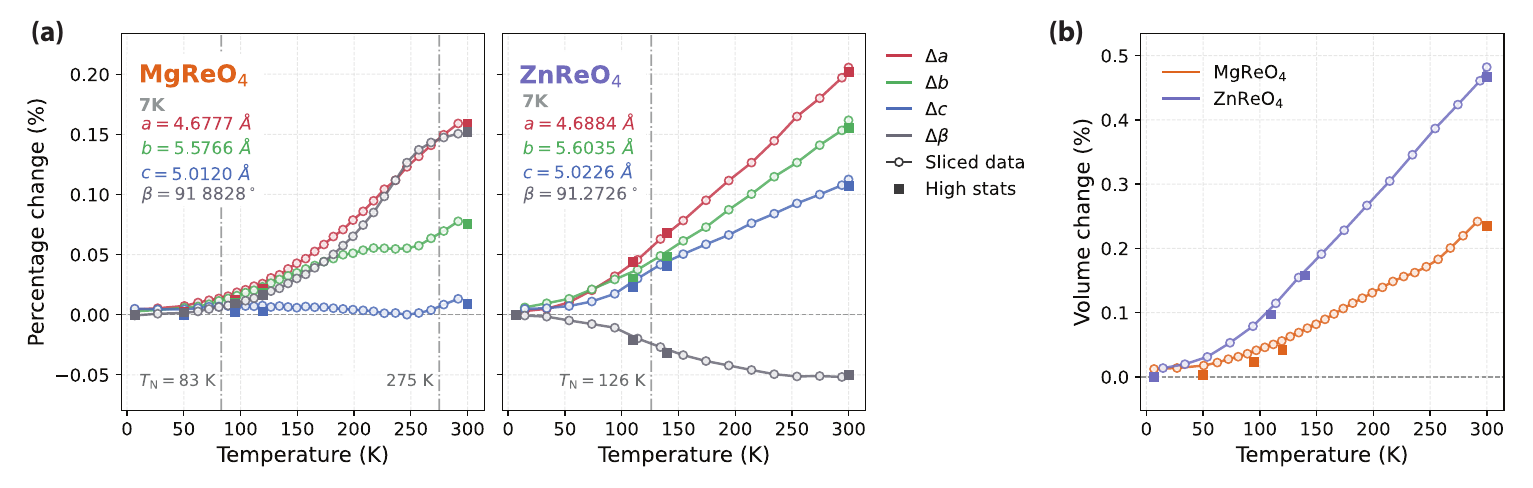}
\caption{Relative changes in (a) the lattice parameters $a$, $b$, and $c$ and monoclinic angle $\beta$, and (b) the unit-cell volume of MgReO$_4$ and ZnReO$_4$. All changes are referenced to the corresponding values obtained from the 7~K high-statistics refinements according to Eq.~(\ref{eq:relative_change}). Open circles and filled squares denote parameters obtained from Rietveld refinements of 20-minute time-sliced data and static high-statistics data, respectively. In (a), the N\'eel temperatures and the previously reported heat-capacity maximum at 275~K associated with the $T^*$ anomaly in MgReO$_4$ are marked.}
 \label{fig:lattice}
\end{figure*}

Figure~\ref{fig:lattice}(a) compares the relative changes in lattice parameters $a$, $b$, $c$, and $\beta$. ZnReO$_4$ exhibits smooth, monotonic expansion of all three lattice parameters, with the hierarchy $\Delta a>\Delta b>\Delta c$. Between 7 and 300~K, $a$ increases by approximately 0.20\%, $b$ by 0.16\%, and $c$ by 0.11\%. The monoclinic angle decreases smoothly and approaches a nearly constant value at the highest temperatures. These curves provide an experimental comparison with MgReO$_4$.

MgReO$_4$ shares the same overall anisotropic hierarchy, but the expansion along $b$ and especially $c$ is substantially smaller than in ZnReO$_4$. $b$ increases up to approximately 200~K, followed by a broad plateau extending to about 250--260~K, before resuming expansion. $c$ is nearly temperature independent up to 200~K, followed by a slight contraction leading to a shallow minimum near 250~K, before beginning to increase. For the BS-bank Rietveld refinements, the mean $c$ over 245--260~K is $(1.75\pm0.55)\times10^{-4}$~\AA\ below the mean over 190--210~K and $(5.05\pm0.60)\times10^{-4}$~\AA\ below the mean over 275--295~K. We propagated the formal refinement uncertainties assuming independent errors between temperature bins. In contrast, $a$ increases monotonically throughout the measured range. 

The evolution of the monoclinic angle provides a further distinction between the compounds. Whereas $\beta$ decreases smoothly in ZnReO$_4$ and starts to saturate above 250~K, it increases in MgReO$_4$. Between 7 and 300~K, $\beta$ increases by $0.14^\circ$, with a more pronounced increase above approximately 180--200~K before approaching saturation
near $T^*$. Together with the suppressed expansion along $b$ and the shallow minimum in $c$, this behavior reveals a change in the shape of the monoclinic unit cell.

The unit-cell volumes of both compounds increase monotonically [Fig.~\ref{fig:lattice}(b)]. In ZnReO$_4$, the volume expansion reflects the increase of all three lattice parameters, with an additional small contribution from the decrease of $\beta$ toward $90^\circ$. In MgReO$_4$, the volume continues to expand through the interval containing the $b$-axis plateau and shallow $c$-axis minimum: the expansion along $a$ and $b$ outweighs the contributions from the slight contraction along $c$ and the increase in $\beta$. Above approximately 250~K, the volume increases more rapidly as $b$ and $c$ begin to expand. 

The data suggest a possible change in slope near $T_{\mathrm N}$ in both compounds. However, the present temperature binning, particularly at low temperatures for MgReO$_4$, does not allow a statistically significant assessment of this feature.

\subsection{Independent profile-fit check}
\begin{figure*}
    \centering
    \includegraphics[width=.8\linewidth]{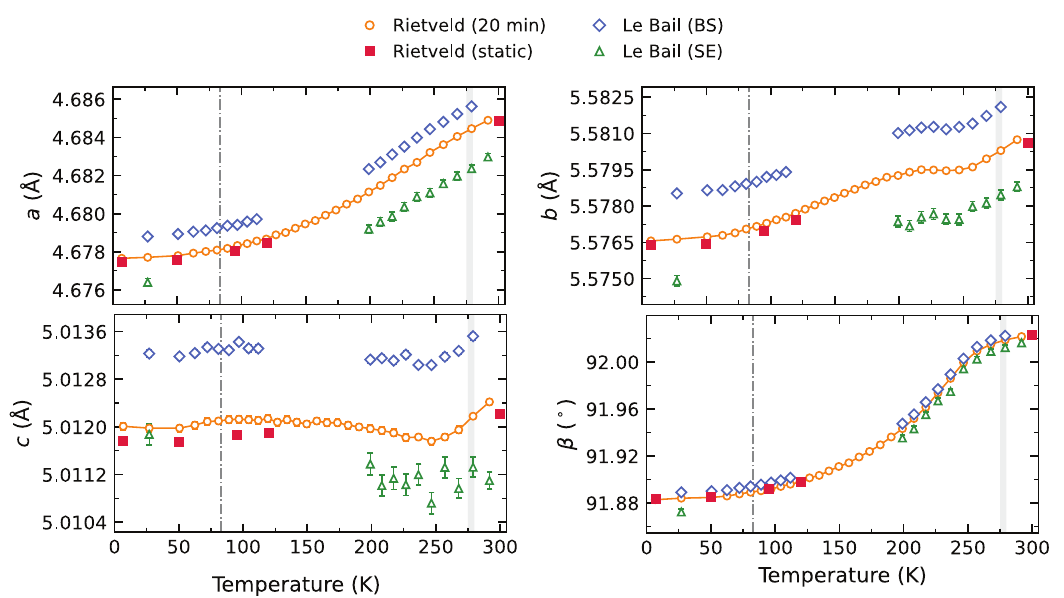}
    \caption{Temperature dependence of the absolute lattice parameters $a$, $b$, and $c$ and monoclinic angle $\beta$ of MgReO$_4$. Rietveld refinements of 20-minute time-sliced data and static high-statistics data are compared with Le Bail fits to the BS-bank and SE-bank data. The vertical markers indicate the magnetic ordering temperature $T_{\mathrm N}=83$~K and the temperature range of the previously reported bulk anomaly, $T^*\simeq275$--280~K. The corresponding relative percentage changes are compared in Fig.~S4 of the SM.}
    \label{fig:lebail}
\end{figure*}

We used Le Bail fits to check whether the MgReO$_4$ lattice anomaly could arise from correlations with refined atomic coordinates and displacement parameters. Figure~\ref{fig:lebail} compares the lattice parameters and monoclinic angle obtained from BS-bank Rietveld refinements and Le Bail fits to both the BS- and SE-bank data. The absolute values differ between analyses and detector banks, most noticeably for $c$, whose temperature-dependent variation is small.

The BS-bank Le Bail fits reproduce the $b$-axis plateau and shallow minimum in $c$ observed in the Rietveld refinements. The relative percentage changes are compared in Fig.~S4 of the SM, where the BS-bank Le Bail and Rietveld curves closely overlap. The SE-bank results show larger scatter, particularly for $c$, and give a less precise comparison. The agreement between the two BS-bank analyses shows that the observed features persist without refining atomic coordinates or displacement parameters.

The static high-statistics measurements also follow the overall lattice trends. Their temperatures do not always coincide with the centers of the time-sliced bins. Nevertheless, these measurements provide an additional check on the structural model and overall lattice evolution. Representative refinement profiles and selected lattice parameters are presented in Fig.~S3 and Table~S3 of the SM.

\section{Discussion}
The anisotropic lattice response of MgReO$_4$ develops over a broad temperature interval rather than at a single sharply defined temperature. The plateau in $b$ and the shallow minimum in $c$ are most apparent between approximately 200 and 260~K, while the renewed increase in $b$ and $c$ and the tendency of $\beta$ to saturate occur closer to the previously reported susceptibility and heat-capacity features at  $T^*\simeq275$--280~K~\cite{Nocerino_MgReO4}. The thermodynamic anomaly therefore lies near the upper end of the broad lattice response, above the $c$-axis minimum near 250~K. This proximity suggests that the lattice response and the bulk features may be connected, but whether they share a microscopic origin remains open. 

We observe no additional Bragg reflections or peak splitting, and the diffraction patterns remain consistent with the average $P2/c$ structure throughout 7--300~K. Within the resolution of the present measurements, we find no evidence for a long-range symmetry-lowering structural transition at $T^*$; however, subtle local distortions or an isostructural rearrangement cannot be excluded.

The contrast between MgReO$_4$ and ZnReO$_4$ helps place this response in context. Strongly anisotropic thermal expansion is common in wolframites, including ZnWO$_4$ and MgWO$_4$~\cite{Trots2009,ZinElabedine2025}. ZnReO$_4$ also expands least along $c$, but all three lattice parameters increase monotonically. What distinguishes MgReO$_4$ is the broad $b$-axis plateau and shallow contraction along $c$, while its unit-cell volume continues to increase. The robustness of the wolframite structure in MgReO$_4$ has also been demonstrated in a recent high-pressure X-ray diffraction study, which found no structural transition up to 20~GPa~\cite{LopezMoreno2026}.

The opposite trends in $\beta$ reinforce the contrast between the two compounds: the angle increases in MgReO$_4$ but decreases in ZnReO$_4$. Alongside the different responses of $b$ and $c$, this points to different evolutions of the monoclinic unit-cell geometry despite the persistence of the same average space group. 

One possible origin is anharmonic lattice dynamics. Transverse or low-energy optical modes can generate strongly anisotropic expansion, including contraction along individual crystallographic directions, while the overall volume expansion remains positive~\cite{Barrera2005,Chatterji2008}. Replacing Mg with Zn changes both atomic mass and local bonding, which can lead to different phonon spectra. Low-energy lattice modes can also contribute substantially to heat capacity, as demonstrated in the negative-thermal-expansion compound ZrW$_2$O$_8$~\cite{Ramirez1998}. Phonon dynamics could therefore contribute to both the heat-capacity feature and the anomalous thermal expansion observed in MgReO$_4$.

A second possibility is a local structural or electronic rearrangement that preserves the average $P2/c$ symmetry. In Ba$_2$MgReO$_6$, quadrupolar ordering accompanies a cubic-to-tetragonal structural transition near 33~K, above the magnetic transition near 18~K~\cite{Hirai2020_multipol,Soh2024}. Spectroscopic studies have also demonstrated dynamic Jahn--Teller effects and spin--orbit--lattice entanglement in related rhenium double perovskites~\cite{Frontini2024,ivkovi2024}. In the wolframites, however, the ReO$_6$ octahedra are already strongly distorted, and our previous calculations found a splitting of approximately 200~meV between the two lowest crystal-field doublets~\cite{ugne_AReO4}. A direct analogue of the cubic-to-tetragonal transition is therefore not expected here. However, local distortions, short-range orbital correlations, or an isostructural rearrangement remain possible.

Finally, magnetoelastic coupling to short-range spin-orbital correlations could contribute to the lattice response. Our previous first-principles calculations found that the strongest exchange interactions connect Re sites in edge-sharing octahedra, with nearest-neighbor interactions playing a more dominant role in MgReO$_4$ than in ZnReO$_4$~\cite{ugne_AReO4}. Short-range correlations can produce broad magnetic heat-capacity maxima without long-range ordering, as illustrated by spin-chain models~\cite{Johnston2000}. Such correlations can also couple strongly to the lattice: in MnTe, a substantial magnetovolume response begins in the paramagnetic state above the ordering temperature~\cite{Baral2023}. Short-range spin-orbital correlations could therefore contribute to the heat-capacity feature in MgReO$_4$ and, through magnetoelastic coupling, to its anomalous thermal expansion.

Further experiments are needed to identify the origin of the anomaly at $T^*$. Inelastic neutron scattering could probe phonons and search for magnetic excitations~\cite{Tennant1995}, while Raman spectroscopy could track changes in lattice vibrations. Total-scattering measurements could reveal local distortions of the ReO$_6$ octahedra~\cite{Senn2016}. Measurements of heat capacity and magnetic susceptibility for both compounds over the same temperature range and at different applied magnetic fields could help test whether the bulk anomaly is of magnetic origin~\cite{Breunig2013}.

\section{Conclusion}

Temperature-dependent NPD shows contrasting thermal expansion in MgReO$_4$ and ZnReO$_4$ between 7 and 300~K. The diffraction patterns of both compounds remain within the average monoclinic $P2/c$ structure, with no evidence of a long-range symmetry-lowering transition within the resolution of the measurements.

ZnReO$_4$ shows smooth, monotonic anisotropic expansion. MgReO$_4$ develops a broad plateau in $b$ and a shallow minimum in $c$ near 250~K, while the unit-cell volume continues to increase monotonically. The plateau and dip are also visible in Le Bail fits to the same BS-bank patterns.

The broader lattice response in MgReO$_4$ extends toward the previously reported bulk features at $T^*\simeq275$--280~K. Anharmonic lattice dynamics, a local or isostructural rearrangement, and magnetoelastic coupling to short-range correlations remain possible microscopic explanations. 

\begin{acknowledgments}
The NPD experiments were performed on the iMATERIA diffractometer (BL20) at the Materials and Life Science Experimental Facility of J-PARC under the user program (Proposal No.~2024A0387). We thank Takafumi Hawai for his support during the NPD measurements. This research was supported by the Swedish Research Council (VR, Grants No.~2021-06157 and No.~2022-03936), the Swedish Foundation for Strategic Research through the Swedish national graduate school in neutron scattering (SwedNess), the Carl Tryggers Foundation for Scientific Research (CTS-22:2374), and the Knut and Alice Wallenberg Foundation (Grant No.~2021.0150). U.M. acknowledges support from the KTH-SCI doctoral excellence program. J.S. acknowledges support from the Japan Society for the Promotion of Science through KAKENHI Grants No.~JP23H01840 and No.~JP24H00042.
\end{acknowledgments}

\section*{Data Availability}
The data and analysis files that support this study are available from the corresponding authors upon reasonable request.

\section*{{Ethics declarations}}
The authors declare no competing interests.\par\smallskip
A large language model (LLM) was used to assist with English-language editing, formatting, and improving the clarity and coherence of the manuscript. All suggested changes were reviewed and verified by the authors, who retain full responsibility for the scientific content, interpretations, and conclusions.

\nocite{Toby_2006}
\bibliographystyle{apsrev4-2}
\bibliography{refs}

\end{document}

\title{\texorpdfstring{Supplemental Material for ``Contrasting anisotropic lattice responses of the $5d^1$ wolframites MgReO$_4$ and ZnReO$_4$''}{Supplemental Material for ``Contrasting anisotropic lattice responses of the 5d1 wolframites MgReO4 and ZnReO4''}}

\author{Ugne~Miniotaite}
\email{ugnem@kth.se}
\affiliation{Department of Applied Physics, KTH Royal Institute of Technology, SE-106 91 Stockholm, Sweden}

\author{Hiroya~Sakurai}
\affiliation{National Institute for Materials Science (NIMS), Namiki, Tsukuba, Ibaraki, Japan}

\author{Jun~Sugiyama}
\affiliation{Neutron Science and Technology Center, Comprehensive Research Organization for Science and Society (CROSS), Tokai, Japan}

\author{Toru~Ishigaki}
\affiliation{Neutron Industrial Application Promotion Center, Comprehensive Research Organization for Science and Society (CROSS), Tokai, Japan}

\author{Martin~M{\aa}nsson}
\email{condmat@kth.se}
\affiliation{Department of Applied Physics, KTH Royal Institute of Technology, SE-106 91 Stockholm, Sweden}

\author{Yasmine~Sassa}
\email{sassa@kth.se}
\affiliation{Department of Applied Physics, KTH Royal Institute of Technology, SE-106 91 Stockholm, Sweden}

\maketitle

\section{\texorpdfstring{Previously reported bulk anomaly in MgReO$_4$}{Previously reported bulk anomaly in MgReO4}}
\begin{figure}[hb]
 \centering
 \includegraphics[width=0.98\linewidth]{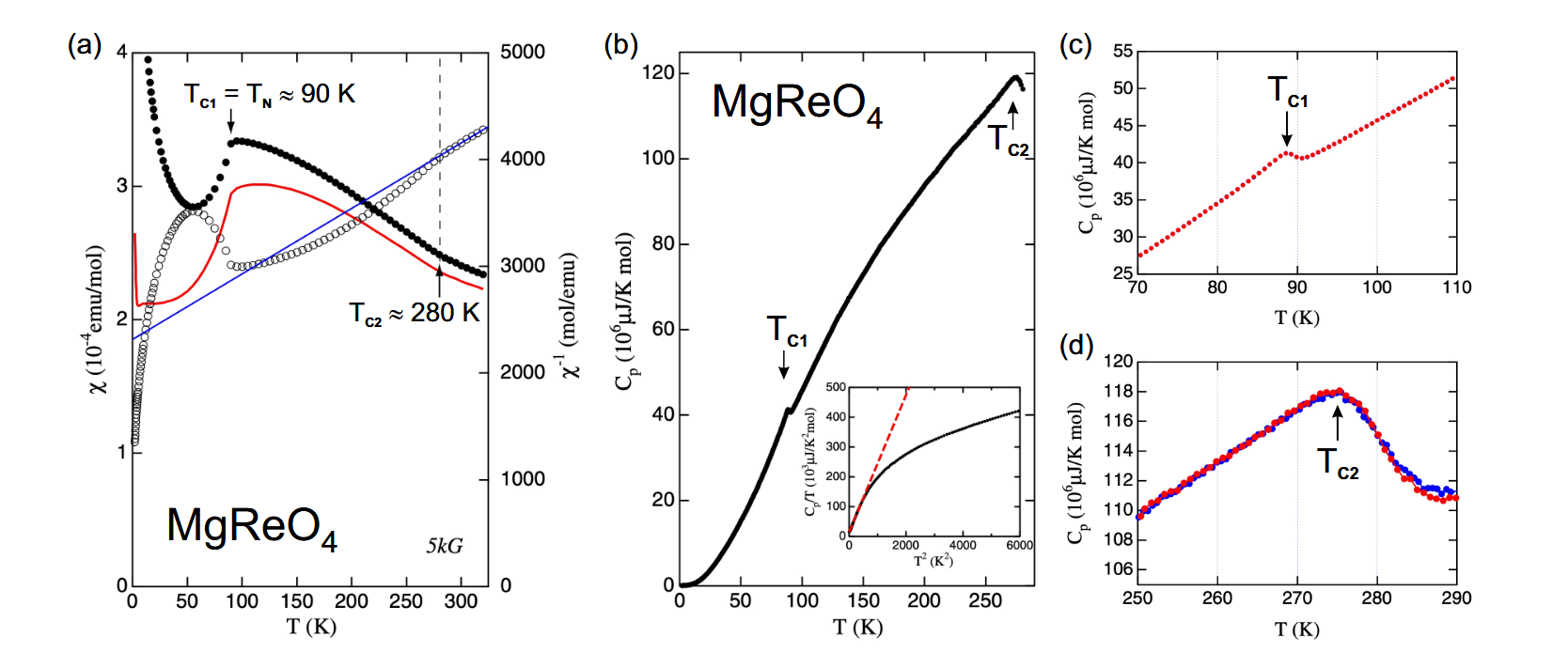}
 \caption{Previously reported bulk response of MgReO$_4$. (a) Corrected dc susceptibility in 5~kG (left axis) and inverse susceptibility (right axis). (b) Heat capacity over the full measured range. (c,d) Expanded views of the heat-capacity features near the magnetic transition and the high-temperature anomaly, respectively. Red and blue points in panel (d) denote heating and cooling. The marker near 90~K identifies the feature in the earlier bulk data,
and the N\'eel temperature $T_{\mathrm N}\simeq83$~K used in the main text comes from weak-transverse-field $\mu^+$SR~\cite{ugne_AReO4}. Adapted from Ref.~\cite{Nocerino_MgReO4}.}
 \label{fig:bulk}
\end{figure}

Figure~\ref{fig:bulk} reproduces the bulk measurements used to define $T^*$. The corrected dc susceptibility measured in a field of 5~kG changes slope near 280~K. The heat capacity contains a weak, reversible feature near 275~K, in addition to the lower-temperature feature associated with magnetic ordering~\cite{Nocerino_MgReO4}. 

\section{Temperature evolution of the diffraction patterns}

\begin{figure}[ht]
 \centering
 \includegraphics[width=0.98\linewidth]{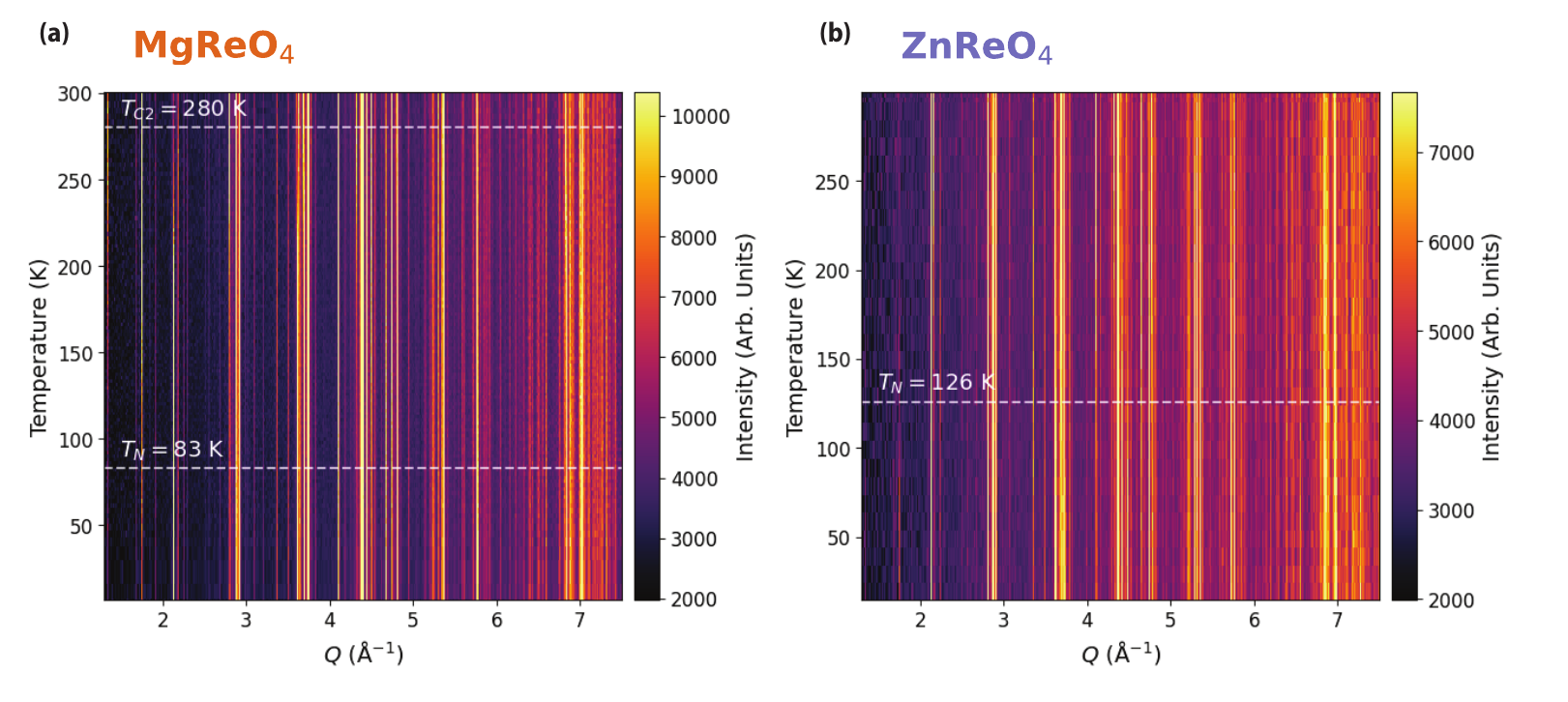}
 \caption{Temperature-dependent BS-bank neutron powder diffraction maps for (a) MgReO$_4$ measured on cooling and (b) ZnReO$_4$ measured on heating. The horizontal lines mark the magnetic ordering temperatures and, for MgReO$_4$, the temperature of the previously reported bulk anomaly. Each panel has its own color normalization.}
 \label{fig:heatmap}
\end{figure}

Figure~\ref{fig:heatmap} shows the time-sliced diffraction patterns from the BS bank. Between 7 and 300~K, we resolve no additional Bragg reflections or peak splitting. The patterns remain consistent with the average monoclinic $P2/c$ structure. Subtle local distortions or other structural changes below the sensitivity of these measurements remain possible.

\section{Temperatures of time-sliced data}

Tables~\ref{tab:temperatures_mg} and~\ref{tab:temperatures_zn} give the average temperature $\langle T\rangle$ and the lowest and highest
temperatures, $T_{\min}$ and $T_{\max}$, for each 20-minute time slice. The average is the arithmetic mean of the temperature-log readings
recorded during that measurement.

\begin{table}[tbp]
\centering
\renewcommand{\baselinestretch}{1}
\small
\caption{Average, minimum, and maximum temperatures for MgReO$_4$ measured on cooling.}
\label{tab:temperatures_mg}
\setlength{\tabcolsep}{8pt}
\renewcommand{\arraystretch}{0.95}
\begin{tabular}{rrr}
\toprule
Average $T$ (K) & $T_{\min}$ (K) & $T_{\max}$ (K) \\
\midrule
291.735 & 284.760 & 299.800 \\
279.005 & 273.370 & 284.730 \\
267.831 & 262.400 & 273.330 \\
257.036 & 251.760 & 262.370 \\
246.594 & 241.490 & 251.730 \\
236.488 & 231.550 & 241.490 \\
226.696 & 221.920 & 231.520 \\
217.213 & 212.570 & 221.890 \\
207.999 & 203.490 & 212.570 \\
199.034 & 194.650 & 203.460 \\
190.300 & 186.030 & 194.630 \\
181.777 & 177.610 & 186.000 \\
173.460 & 169.370 & 177.580 \\
165.330 & 161.310 & 169.350 \\
157.364 & 153.440 & 161.310 \\
149.544 & 145.680 & 153.410 \\
141.858 & 138.060 & 145.680 \\
134.275 & 130.540 & 138.060 \\
126.766 & 123.040 & 130.510 \\
119.309 & 115.620 & 123.040 \\
111.869 & 108.140 & 115.590 \\
104.388 & 100.640 & 108.140 \\
96.802 & 92.952 & 100.620 \\
89.027 & 85.045 & 92.952 \\
80.907 & 76.675 & 85.045 \\
72.223 & 67.614 & 76.675 \\
62.517 & 57.101 & 67.583 \\
50.575 & 43.071 & 57.101 \\
27.030 & 6.785 & 43.014 \\
6.784 & 6.668 & 6.922 \\
\bottomrule
\end{tabular}
\end{table}

\begin{table}[tbp]
\centering
\renewcommand{\baselinestretch}{1}
\small
\caption{Average, minimum, and maximum temperatures for ZnReO$_4$ measured on heating.}
\label{tab:temperatures_zn}
\setlength{\tabcolsep}{8pt}
\renewcommand{\arraystretch}{0.95}
\begin{tabular}{rrr}
\toprule
Average $T$ (K) & $T_{\min}$ (K) & $T_{\max}$ (K) \\
\midrule
14.656 & 6.614 & 24.566 \\
33.971 & 24.623 & 43.871 \\
53.700 & 43.871 & 63.614 \\
73.725 & 63.614 & 83.831 \\
93.903 & 83.831 & 103.960 \\
114.048 & 104.020 & 124.100 \\
134.162 & 124.100 & 144.210 \\
154.232 & 144.210 & 164.220 \\
174.284 & 164.280 & 184.320 \\
194.334 & 184.320 & 204.350 \\
214.383 & 204.410 & 224.390 \\
234.406 & 224.390 & 244.390 \\
254.423 & 244.450 & 264.420 \\
274.454 & 264.420 & 284.470 \\
294.009 & 284.530 & 300.350 \\
299.873 & 299.830 & 299.940 \\
\bottomrule
\end{tabular}
\end{table}

The lowest-temperature MgReO$_4$ measurement has an average temperature of 6.784~K and covers 6.668--6.922~K. The next time slice has an average of 27.030~K and covers 6.785--43.014~K. Near $T_{\mathrm N}\simeq83$~K, the measurements cover 76.675--85.045~K and 85.045--92.952~K. These ranges limit the resolution of a narrow feature at the magnetic ordering temperature. No clear lattice anomaly is resolved at $T_{\mathrm N}$. The ZnReO$_4$ table lists the 16 time slices used in the refinement series.

\section{High-statistics and time-sliced refinements}
\begin{figure}[ht]
 \centering
 \includegraphics[width=0.98\linewidth]{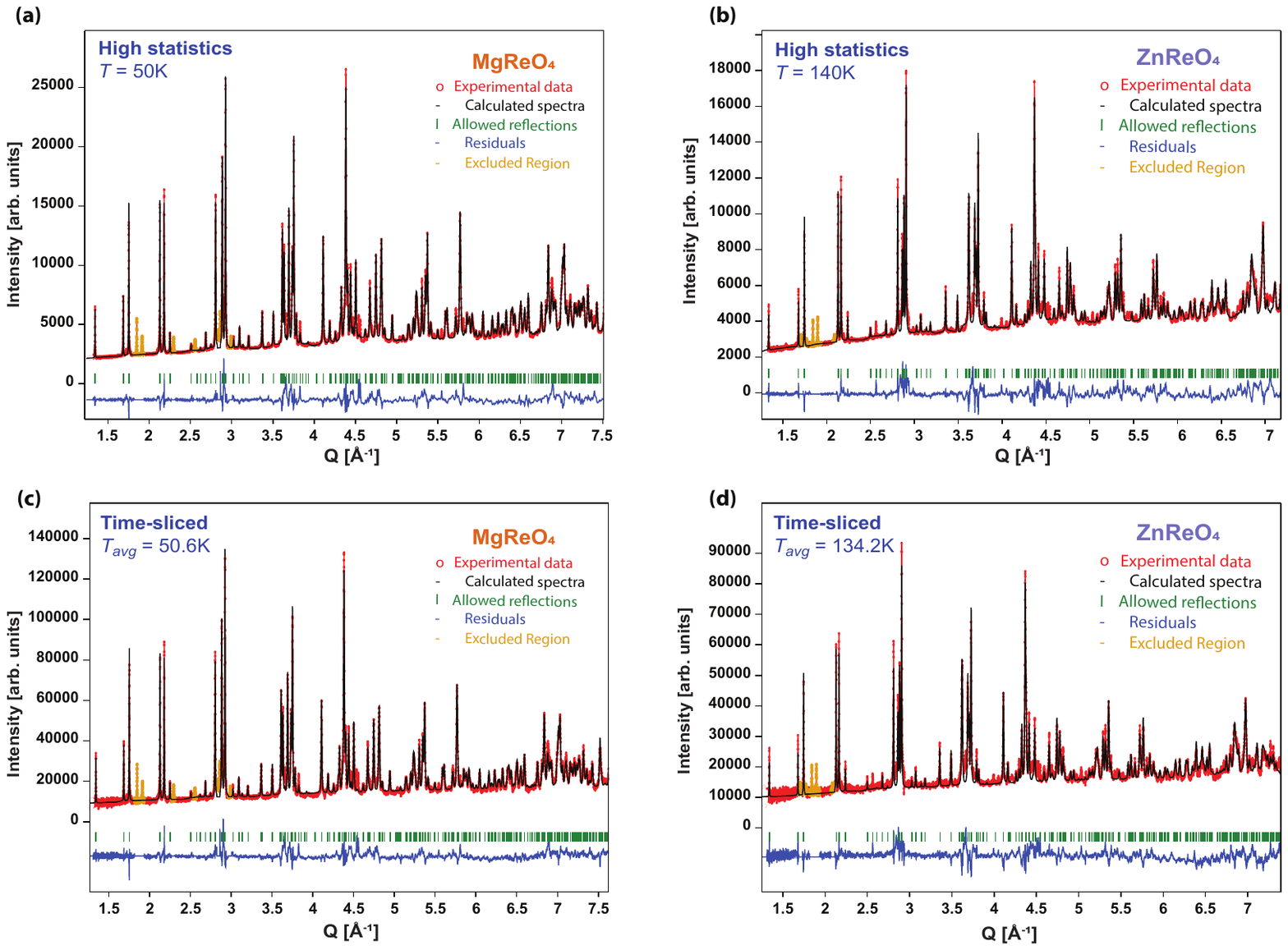}
 \caption{Representative BS-bank Rietveld refinements. The upper panels show high-statistics (HS) measurements of MgReO$_4$ at 50~K (left) and ZnReO$_4$ at 140~K (right). The lower panels show 20-minute time-sliced data with mean temperatures of 50.575~K for MgReO$_4$ (left; 43.071--57.101~K) and 134.162~K for ZnReO$_4$ (right; 124.10--144.21~K). Observed and calculated intensities, their difference, allowed $P2/c$ reflections, and excluded intervals are indicated in each panel. The four intensity scales are independent.}
 \label{fig:refinements}
\end{figure}

The 7~K high-statistics (HS) Rietveld refinements of MgReO$_4$ and ZnReO$_4$ are reported in Ref.~\cite{ugne_AReO4}. We use the lattice parameters and
unit-cell volumes from those refinements (Table~S1 of that reference) to calculate the relative changes shown in the main text. Additional static HS patterns were accumulated for approximately 11.4~h (285 million counts) at 50~K for MgReO$_4$ and 15.8~h (395 million counts) at 140~K for ZnReO$_4$. Figure~\ref{fig:refinements} compares these data with representative 20-minute time-sliced (TS) patterns. The representative diffraction patterns were refined using the monoclinic $P2/c$ structural model. 

Table~\ref{tab:metrics} lists selected parameters obtained from the HS and TS refinements, while Table~\ref{tab:fitquality} summarizes the corresponding fit statistics.  

\begin{table*}[t]
\centering
\caption{Lattice parameters from representative BS-bank Rietveld refinements of high-statistics (HS) measurements and 20-minute time slices (TS). Parentheses give the standard uncertainties scaled by $\sqrt{\chi^2}$ in FullProf~\cite{RodriguezCarvajal1993}.}
\label{tab:metrics}
\footnotesize
\setlength{\tabcolsep}{2.5pt}
\begin{tabular}{llrrrrrr}
\toprule
Compound & Acquisition & $\langle T\rangle$ (K) & $a$ (\AA) & $b$ (\AA) &
$c$ (\AA) & $\beta$ (deg) & $V$ (\AA$^3$) \\
\midrule
MgReO$_4$ & HS & 50.0 & 4.6776(1) & 5.5764(1) & 5.0118(1) & 91.885(1) & 130.657(3) \\
MgReO$_4$ & TS & 50.575 & 4.677804(61) & 5.576733(75) & 5.011978(62) & 91.8847(11) & 130.676(3) \\
ZnReO$_4$ & HS & 140.0 & 4.6916(1) & 5.6062(1) & 5.0247(1) & 91.243(1) & 132.129(4) \\
ZnReO$_4$ & TS & 134.162 & 4.691401(92) & 5.60626(11) & 5.024717(85) & 91.2480(15) & 132.125(4) \\
\bottomrule
\end{tabular}
\end{table*}

\begin{table*}[t]
\centering
\renewcommand{\baselinestretch}{1}
\caption{Fit statistics for the representative BS-bank Rietveld refinements. The $R$ factors include the background and were calculated using all non-excluded profile points. The reduced chi-squared is $\chi^2=(R_{\mathrm{wp}}/R_{\mathrm{exp}})^2$, and the goodness-of-fit parameter is $S=R_{\mathrm{wp}}/R_{\mathrm{exp}}=\sqrt{\chi^2}$.}
\label{tab:fitquality}
\small
\setlength{\tabcolsep}{4pt}
\begin{tabular}{llrrrrr}
\toprule
Compound & Acquisition & $R_p$ (\%) & $R_{\mathrm{wp}}$ (\%) &
$R_{\mathrm{exp}}$ (\%) & $\chi^2$ & $S$ \\
\midrule
MgReO$_4$ & HS, 50 K & 3.88 & 6.37 & 0.296 & 462 & 21.5 \\
MgReO$_4$ & TS, 50.575 K & 5.98 & 8.47 & 2.27 & 13.9 & 3.73 \\
ZnReO$_4$ & HS, 140 K & 3.63 & 5.42 & 0.606 & 80.0 & 8.94 \\
ZnReO$_4$ & TS, 134.162 K & 6.19 & 8.06 & 2.33 & 12.0 & 3.46 \\
\bottomrule
\end{tabular}
\end{table*}

The high count totals of the HS measurements produce small $R_{\mathrm{exp}}$ values, making the remaining profile mismatch more apparent in $\chi^2$~\cite{Toby_2006}. The HS and TS measurements were acquired at different temperatures, and some differences between the refined lattice parameters exceed the reported refinement uncertainties. The BS-bank Rietveld and Le Bail fits both give their lowest sampled
$c$ near 250~K and show a similar increase at higher temperatures. Using the same temperature bins for both methods, the drop in $c$ from the lower-temperature bins to the minimum is smaller in the Le Bail fit and falls within its formal uncertainty. The estimated dip depth is therefore sensitive to the refinement method.

\begin{table*}[t]
\centering
\renewcommand{\baselinestretch}{1}
\small
\caption{Differences between the TS and HS lattice metrics,
$\Delta x=x_{\mathrm{TS}}-x_{\mathrm{HS}}$. The combined formal uncertainty is $\sigma_{\mathrm{comb}}=(\sigma_{\mathrm{TS}}^2+ \sigma_{\mathrm{HS}}^2)^{1/2}$, assuming independent fit errors. The standardized difference is $z=\Delta x/\sigma_{\mathrm{comb}}$. $z$ is calculated before rounding the values for this table.
Cell differences use the final parameter lists; volume differences
use the printed volumes and their uncertainties. These diagnostics
do not account for temperature differences or systematic errors.}
\label{tab:comparison}
\begin{tabular}{llrrr}
\toprule
Compound & Metric & $\Delta x$ & $\sigma_{\mathrm{comb}}$ & $z$ \\
\midrule
MgReO$_4$ & $a$ (\AA) & +0.0002236 & 0.0000874 & +2.56 \\
MgReO$_4$ & $b$ (\AA) & +0.0003061 & 0.0001082 & +2.83 \\
MgReO$_4$ & $c$ (\AA) & +0.0002241 & 0.0000882 & +2.54 \\
MgReO$_4$ & $\beta$ (deg) & -0.000435 & 0.001610 & -0.27 \\
MgReO$_4$ & $V$ (\AA$^3$) & +0.019000 & 0.004243 & +4.48 \\
ZnReO$_4$ & $a$ (\AA) & -0.0002179 & 0.0001264 & -1.72 \\
ZnReO$_4$ & $b$ (\AA) & +0.0000329 & 0.0001504 & +0.22 \\
ZnReO$_4$ & $c$ (\AA) & +0.0000529 & 0.0001179 & +0.45 \\
ZnReO$_4$ & $\beta$ (deg) & +0.004517 & 0.002028 & +2.23 \\
ZnReO$_4$ & $V$ (\AA$^3$) & -0.004000 & 0.005657 & -0.71 \\
\bottomrule
\end{tabular}
\end{table*}

\section{Refinement functions, ranges, and parameter conventions}

The BS-bank Rietveld refinements used a single monoclinic $P2/c$ structural model. Impurity
fractions could not be determined \cite{ugne_AReO4}. The TOF profile used a pseudo-Voigt function convolved with back-to-back exponential decay functions
(FullProf option \texttt{Npr=9}). The background was obtained by linear interpolation between refined background points. Regions containing impurity peaks were excluded from the refinements; these regions are listed in Table~\ref{tab:excluded}. The numerical fit-quality parameters for the HS and TS refinements are given in Table~\ref{tab:fitquality}.

\begin{table*}[t]
\centering
\renewcommand{\baselinestretch}{1}
\small
\caption{Fitting ranges in $d$ and $Q$. The SE row refers to
the Le Bail fit.}
\label{tab:settings}
\setlength{\tabcolsep}{5pt}
\begin{tabular}{lrr}
\toprule
Fit & $d$ (\AA) & $Q$ (\AA$^{-1}$)  \\
\midrule
MgReO$_4$ BS, TS & 0.82197--4.79826 & 1.30947--7.64403 \\
MgReO$_4$ BS, HS & 0.82197--4.79826 & 1.30947--7.64403  \\
ZnReO$_4$ BS, TS & 0.84662--4.79824 & 1.30948--7.42152  \\
ZnReO$_4$ BS, HS & 0.84661--4.79848 & 1.30941--7.42161  \\
MgReO$_4$ SE, Le Bail & 1.25844--6.46472 & 0.97192--4.99283  \\
\bottomrule
\end{tabular}
\end{table*}

\begin{table*}[t]
\centering
\renewcommand{\baselinestretch}{1}
\small
\caption{Excluded regions in $Q$ (\AA$^{-1}$), calculated with
the stored TOF calibrations. Interior intervals remove impurity peaks.
One-sided cutoffs apply within each dataset's measured range.
Mg BS values apply to HS and TS; Zn BS values are listed separately
because their calibrations differ. SE values refer to the representative
Le Bail control file.}
\label{tab:excluded}
\setlength{\tabcolsep}{5pt}
\begin{tabular}{rrrr}
\toprule
MgReO$_4$ BS & ZnReO$_4$ BS, TS & ZnReO$_4$ BS, HS & MgReO$_4$ SE \\
\midrule
$Q\geq 7.6440$ & $Q\geq 7.4215$ & $Q\geq 7.4216$ & $Q\geq 4.9928$ \\
2.9676--3.0077 & 2.0705--2.1122 & 2.0705--2.1122 & 2.2718--2.3134 \\
2.8263--2.8626 & 1.8170--1.9355 & 1.8170--1.9354 & 1.8045--1.9737 \\
2.5436--2.5730 & 1.7630--1.7807 & 1.7630--1.7806 & 1.7070--1.7351 \\
2.2711--2.3124 & 1.6895--1.7322 & 1.6894--1.7321 & $Q\leq 0.9719$ \\
1.8170--1.9355 & $Q\leq 1.3095$ & $Q\leq 1.3094$ & --- \\
$Q\leq 1.3095$ & --- & --- & --- \\
\bottomrule
\end{tabular}
\end{table*}

\section{Le Bail profile fits}

Le Bail fits were performed for MgReO$_4$ using the BS and SE detector banks to check the unit-cell metrics without refining atomic coordinates or isotropic displacement parameters. The BS fits used the same $d$ and $Q$ ranges as the BS Rietveld refinements (Table~\ref{tab:settings}). The representative SE fit covered $d=1.2584$--$6.4647$~\AA\ ($Q=0.9719$--$4.9928$~\AA$^{-1}$).

\begin{figure}[ht]
 \centering
 \includegraphics[width=0.82\linewidth]{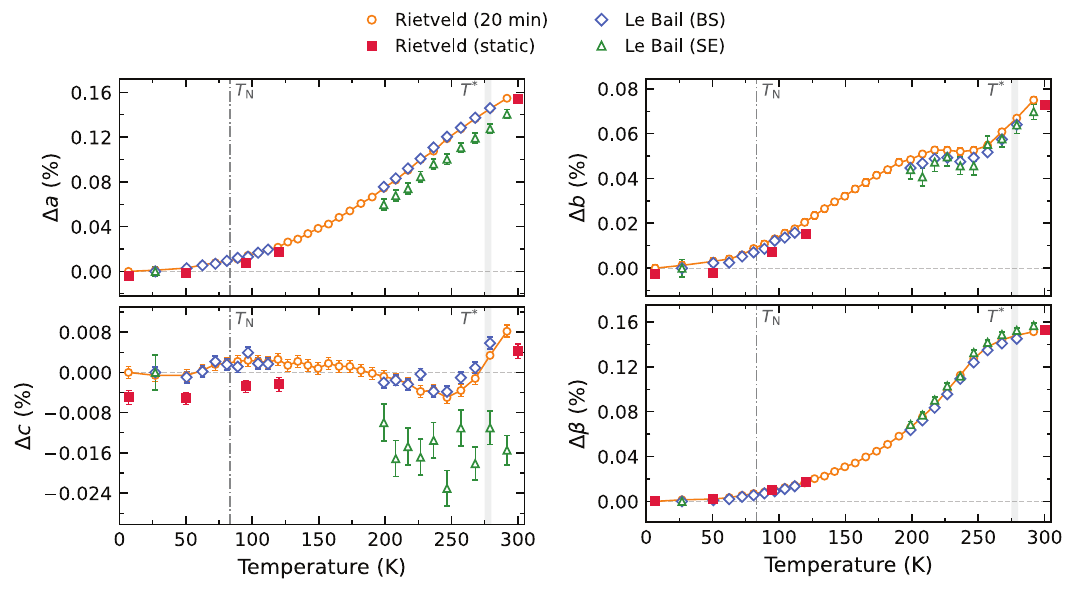}
 \caption{Percentage changes in the MgReO$_4$ lattice parameters and monoclinic angle from BS- and SE-bank Le Bail fits, compared
 with the Rietveld-refined time-sliced series and HS points. Each series is referenced to its own low-temperature value.}
 \label{fig:lebailrelative}
\end{figure}

The BS-bank Le Bail fits reproduce the $b$-axis plateau and shallow minimum in $c$ observed in the Rietveld refinements. 
Figure~\ref{fig:lebailrelative} compares the percentage changes from the Le Bail and Rietveld fits. The SE-bank results show greater scatter, particularly for $c$.

\clearpage
\bibliographystyle{apsrev4-2}
\bibliography{refs}